\documentclass[apj]{emulateapj}

\usepackage{amsmath}
\usepackage[usenames,dvipsnames]{color}
\usepackage{amssymb}

\begin{document}

\title{Upper limits to the magnetic field in central stars of planetary nebulae}
\author{A. Asensio Ramos\altaffilmark{1,2}, M. J. Mart\'\i nez Gonz\'alez\altaffilmark{1,2}, R. Manso Sainz\altaffilmark{1,2}, R. L. M. Corradi\altaffilmark{1,2}, F. Leone\altaffilmark{3}}
\altaffiltext{1}{Instituto de Astrof\'{\i}sica de Canarias, 38205, La Laguna, Tenerife, Spain}
\altaffiltext{2}{Departamento de Astrof\'{\i}sica, Universidad de La Laguna, E-38205 La Laguna, Tenerife, Spain}
\altaffiltext{3}{Dipartimento di Fisica e Astronomia, Universit\'a di Catania, Sezione Astrofisica, Via S. Sofia 78, I-9512, Catania, Italy}
\email{aasensio@iac.es}

\begin{abstract}
More than about twenty central stars of planetary nebulae (CSPN) have been observed
spectropolarimetrically, yet no clear, unambiguous signal of the presence of 
a magnetic field in these objects has been found. 
We perform a statistical (Bayesian) analysis of all the available spectropolarimetric observations 
of CSPN to constrain the magnetic fields on these objects.
Assuming that the stellar field is dipolar and that the dipole axis of the objects are oriented randomly (isotropically),
we find that the dipole magnetic field strength is smaller than 400~G with 95\% probability using
all available observations.
The analysis introduced allows integration of future observations to further constrain 
the parameters of the distribution, and it is general, so that it can be 
easily applied to other classes of magnetic objects.
We propose several ways to improve the upper limits found here.
\end{abstract}

\keywords{magnetic fields --- polarization --- techniques: polarimetric --- methods: data analysis, statistical}
\maketitle

\section{Introduction}
The detection of magnetic fields in the CSPN
has been a subject of considerable interest recently. 
The presence of magnetic fields on CSPN could shed some light on the 
magnetic field of planetary nebulae (PNe) themselves, and hence on the 
role of magnetic fields on the shaping of PNe \citep[][]{ChevalierLuo94, 
Tweedy+95, Guille99, Blackman+01a, Blackman+01, BalickFrank02, Soker04, 
Soker06, Vlemmings+06, Sabin+07}.
Yet, magnetic fields on CSPN have proved elusive 
and so far, no clear, unambiguous spectropolarimetric detection of a magnetic field
in a CSPN has been possible 
\citep{Leone+11, Bagnulo+12, Jordan+12}.

Spectropolarimetry of CSPN is challenging because they are intrinsically faint 
and must be observed at relatively low spectral resolutions, 
which leads to cancellations attenuating the, already weak, polarimetric signals.
Additionally, CSPN present a relatively low number of spectral lines, 
which hampers {\em line addition} techniques that exploit the collective 
contribution of hundreds (or even thousands) of spectral lines to increase 
the signal-to-noise ratio \citep{Semel96, Donati+97, Martinez+08}.

Here we follow a different approach. We aim at constraining the magnetic field 
on CSPN statistically, combining the overall information of all available 
observations to constrain the magnetism of CSPN \citep[in a way similar, but more general, to the
analysis of magnetic fields in RR Lyrae of][]{kolenberg_bagnulo09}. 
A common procedure would be combining the inferred values of the field strengths 
(obtained for example, from least-squares fitting) in a histogram.
However, this is unsuitable because on the one hand, noise introduces large
uncertainties and degeneracies in the determination of the field that are not
properly propagated when carrying out a histogram; and on the other, the 
magnetic field is not a directly measurable quantity, it is {\em inferred} 
from observations. Consequently, a robust inference of the distribution of 
magnetic fields in CSPN is better done within the Bayesian formalism 
\citep[see][2011 for a similar approach]{Bovy+11}.
In fact, we will follow a hierarchical Bayesian approach similar to that recently 
follower by \cite{Hogg+10} to estimate the distribution of eccentricities 
in the orbits of binary stars and exoplanets; but unlike them, 
here, we carry out the full hierarchical Bayesian analysis.

\section{Magnetic field inference under a Bayesian hierarchical analysis}
In magnetized atmospheres, spectral lines show a characteristic
circular polarization pattern dominated by the Zeeman effect.
If the Zeeman splitting is sufficiently small so that it does not
dominate the broadening of the spectral lines, and assuming 
that the magnetic field is roughly constant along the line of sight (LOS) and
has a dipolar topology in the stellar surface, 
this circular polarization flux pattern $F_V$ can be simply modeled as 
\citep[see][]{Landi92,Martinez+12}
\begin{equation}\label{eq01}
F_V(\lambda)=-\alpha B_\parallel \frac{dF_I(\lambda)}{d\lambda},
\end{equation}
where $F_I$ is the intensity flux, and $B_\parallel=B_d \cos \theta_d$,
with $B_d$ being the magnetic field strength in the pole and $\theta_d$ the inclination of the dipole
axis with respect to the LOS. Additionally, 
$\alpha = 1.17\times 10^{-13}\lambda_0^2 g_\mathrm{eff}$ in the absence of limb darkening \citep[see][for the general expression]{Martinez+12}, 
$\lambda_0$ is the central wavelength of the spectral line (in \AA),
and $g_\mathrm{eff}$ is the effective Land\'e factor of the transition \citep{Landi82}.
For a general discussion about the weak-field approximation in stellar
magnetism, we refer to \cite{Martinez+12}. From the observational
point of view, $F_V/F_I$ is easier to measure because it is less prone to errors \citep[e.g.,][]{bagnulo09}. Consequently,
we work instead with 
\begin{equation}\label{eq01b}
\frac{{F}_V(\lambda)}{{F}_I(\lambda)}=-\alpha B_\parallel \frac{1}{{F}_I(\lambda)}\frac{d{F}_I(\lambda)}{d\lambda}.
\end{equation}
Although all the subsequent formalism is presented in terms of ${F}_V(\lambda)$
to simplify the notation, they are still valid provided one substitutes ${F}_V(\lambda)$ for
${F}_V(\lambda)/{F}_I(\lambda)$ and $d{F}_I(\lambda)/d\lambda$ for $(d{F}_I(\lambda)/d\lambda) / {F}_I(\lambda)$.

A number of different methods have been devised to infer the magnetic field 
$B_\parallel$ of an object from a set of observations $D=\{{F}_I(\lambda_1),  
\dots, {F}_I(\lambda_M), {F}_V(\lambda_1), \dots, {F}_I(\lambda_M)\}$ 
of the flux of the Stokes $I$ and Stokes $V$ spectra at some given wavelengths 
$\lambda_1, \dots, \lambda_M$.
Maximum likelihood methods consist on choosing the parameter(s) that maximizes 
the {\em likelihood} function ${\cal L}_\theta$, which is the probability of observing 
the data (here, the intensity and polarization profiles), given the physical 
parameters $\theta$ of the model (here, the average longitudinal magnetic field)
\citep[e.g.,][]{Jaynes03, Martinez+12}.
The likelihood ensues simply from equation~(\ref{eq01}) 
and our model for the noise (here, Gaussian; see Appendix):
\begin{multline}\label{eq02}
{\cal L}_{B_\parallel}=
\left[\prod_{s=1}^S \frac{1}{[2\pi( \sigma_V^2 + \alpha_s^2 B_\parallel^2 \sigma_{I'}^2)]^{(L_s/2)}}\right] 
\\
\times\exp\left\{ -\sum_{s=1}^S\frac{c_1(B_\parallel-\hat{b}_\parallel)^2-c_0}{2(\sigma_V^2+\alpha_s^2 B_\parallel^2 \sigma_{I'}^2)} \right\},
\end{multline}
where $s$ extends over all the $S$ spectral lines in the observed spectrum 
(each with $L_s$ wavelengths),  
$\sigma_V^2$ and $\sigma_{I'}^2$ are the variances of the Gaussian noise
of the circular polarization profile and the derivative of the intensity respectively, and $c_{0, 1}$ and $\hat{b}_\parallel$
are combination of averages of the observed $F_V$ and $F'_I$ defined in the Appendix.
The denominator of each term in the exponential can be understood 
as (twice) the variance of the linear combination $F_V+\alpha B_\parallel F_I'$
of the two random variables $F_V$ and $F_I'$:
$\mathrm{Var}(F_V+\alpha B_\parallel F_I')=
\mathrm{Var}(F_V)+\alpha^2 B^2_\parallel \mathrm{Var}(F_I') + 2 \alpha B_\parallel \mathrm{Cov}(F_V,F_I')=
\sigma_V^2+\alpha^2 B^2_\parallel\sigma^2_{I'}+2\alpha B_\parallel \sigma_V \sigma_{I'} \rho$, assuming that 
$\rho$ is the average correlation between $F_V$ and $F_{I'}$ \citep{Press+92, AsensioManso11}.
For typical modulation schemes, $\sigma^2_I\sim\sigma^2_V$.
Hence, $\sigma^2_{I'}\sim 2\sigma^2_V R^2 /\lambda^2$, where $R$ is the spectral 
resolution and $\lambda$ the characteristic wavelength.
In the optical ($\lambda\approx 5000$ \AA), for $R=3500$, and for 
relatively weak fields, the second and third terms can be neglected so that
equation~(\ref{eq02}) simplifies to
\begin{equation}\label{eq03}
{\cal L}_{B_\parallel}\approx
\left[\prod_{s=1}^S
\frac{1}{[2\pi\sigma_V^2]^{(L_s/2)}}
\right]
\exp\left\{ -C_1(B_\parallel-\hat{B}_\parallel)^2+C_0 \right\},
\end{equation}
where $C_{0, 1}$ and $\hat{B}_\parallel$ have been redefined as in equation~(\ref{eqa8})
(for clarity, the explicit dependence of $\sigma_V^2$ on the spectral line is not shown).
Note that $\hat{B}_\parallel$ is the maximum likelihood estimate for $B_\parallel$ 
\citep[see][]{Martinez+12,Leone+14}.

In a Bayesian analysis, the {\em a posteriori} probability for the presence of
a magnetic field $B_\parallel$ in the object given some observations $D$ 
is simply obtained from the Bayes theorem \citep[e.g.,][]{Jaynes03, Gregory05}
\begin{equation}\label{eq04}
p(B_\parallel|D)=\frac{p(D|B_\parallel) p(B_\parallel)}{p(D)},
\end{equation}
where $p(D|B_\parallel)\equiv {\cal L}_{B_\parallel}$, 
$p(B_\parallel)$ measures our {\em a priori} knowledge on the value of the field, 
and $p(D)$ is the so-called {\em evidence}, which we will not consider explicitly 
in the following since it is just a scaling factor so that the posterior is normalized
($p(D)=\int p(D|B_\parallel)p(B_\parallel) \mathrm{d}B_\parallel$).
There is, in principle, no {\em a priori} reason for the field pointing towards us or away from us,
and it seems likely that the most probable value of $B_\parallel$ be zero.
Hence, a natural choice for $p(B_\parallel)$ would be a normal 
distribution ${\cal N}(0, b^2)$ for some fixed variance $b^2$ (which can be potentially very large).
Equation~(\ref{eq04}) may thus be used to infer the magnetic field of an object
from polarimetric observations 
\citep[see, for example,][]{Asensio+07, Asensio09, Asensio11}.
Instead, here,
we are interested on the statistical distribution of all of them and how this
distribution is constrained by the observations.


\begin{table}
\begin{center}
\caption{Statistical parameters of CSPN\label{tab:table}}
\begin{tabular}{lcccc}
\tableline\tableline
Object (PNe) & $S$ & $\hat{B}_\parallel$ [G]\tablenotemark{a} & $\hat{B}_\parallel$ [G] (with error)\tablenotemark{b} & $C_1$ \\ 
\tableline
NGC 2392 &            4 &   $-1438$ & $-1000 \pm 2000$ & 1.3$\times$10$^{-7}$ \\
NGC 1360 &           12 &   $-1423$ & $-1000 \pm 4000$ & 3.2$\times$10$^{-8}$ \\
NGC 1360 &           16 &   $420$   & $0 \pm 4000$ & 2.9$\times$10$^{-8}$ \\
NGC 1360 &            4 &   $-172$  & $-170 \pm 80$ & 8.2$\times$10$^{-7}$ \\
NGC 6826 &           12 &   $2640$  & $3000 \pm 16000$ & 2.1$\times$10$^{-7}$ \\
NGC 6826 &            4 &   $-6910$ & $-7000 \pm 13000$ & 3.2$\times$10$^{-9}$ \\
PHL 932 &           20 &    $-1238$ & $-1000 \pm 5000$ & 2.3$\times$10$^{-8}$ \\
LSS 1362 &            4 &   $-3560$ & $-3600 \pm 1700$ & 1.6$\times$10$^{-7}$ \\
Abell 36 &            8 &   $208$   & $200 \pm 1200$ & 3.8$\times$10$^{-7}$ \\
Abell 36 &           12 &   $950$   & $900 \pm 1200$ & 4.4$\times$10$^{-7}$ \\
IC 4637 &            4 &    $-1668$ & $-1700 \pm 1400$ & 2.8$\times$10$^{-7}$ \\
LSE 125 &           12 &    $-50$   & $0 \pm 600$ & 1.4$\times$10$^{-6}$ \\
LSE 125 &            4 &    $3381$  & $3400 \pm 1600$ & 2.0$\times$10$^{-7}$ \\
NGC 4361 &            7 &   $-2607$ & $-2600 \pm 1600$ & 2.0$\times$10$^{-7}$ \\
NGC 7293 &            2 &  $-27985$ & $-28000 \pm 11000$ & 4.1$\times$10$^{-9}$ \\
Tc 1 &            6 &       $223$   & $200 \pm 1300$ & 2.9$\times$10$^{-7}$ \\
Tc 1 &            1 &       $7373$  & $7000 \pm 7000$ & 1.9$\times$10$^{-8}$ \\
NGC 6026 &           30 &   $882$   & $900 \pm 500$ & 1.8$\times$10$^{-6}$ \\
HD 44179 &           16 &   $-289$  & $-290 \pm 150$ & 2.3$\times$10$^{-5}$ \\
\tableline
\end{tabular}
\tablenotetext{1}{Maximum likelihood estimate of the longitudinal component of the magnetic field.}
\tablenotetext{2}{Estimation of the field with error bars and taking into account the significant figures.}
\end{center}
\end{table}

\begin{figure}
\includegraphics[width=\columnwidth]{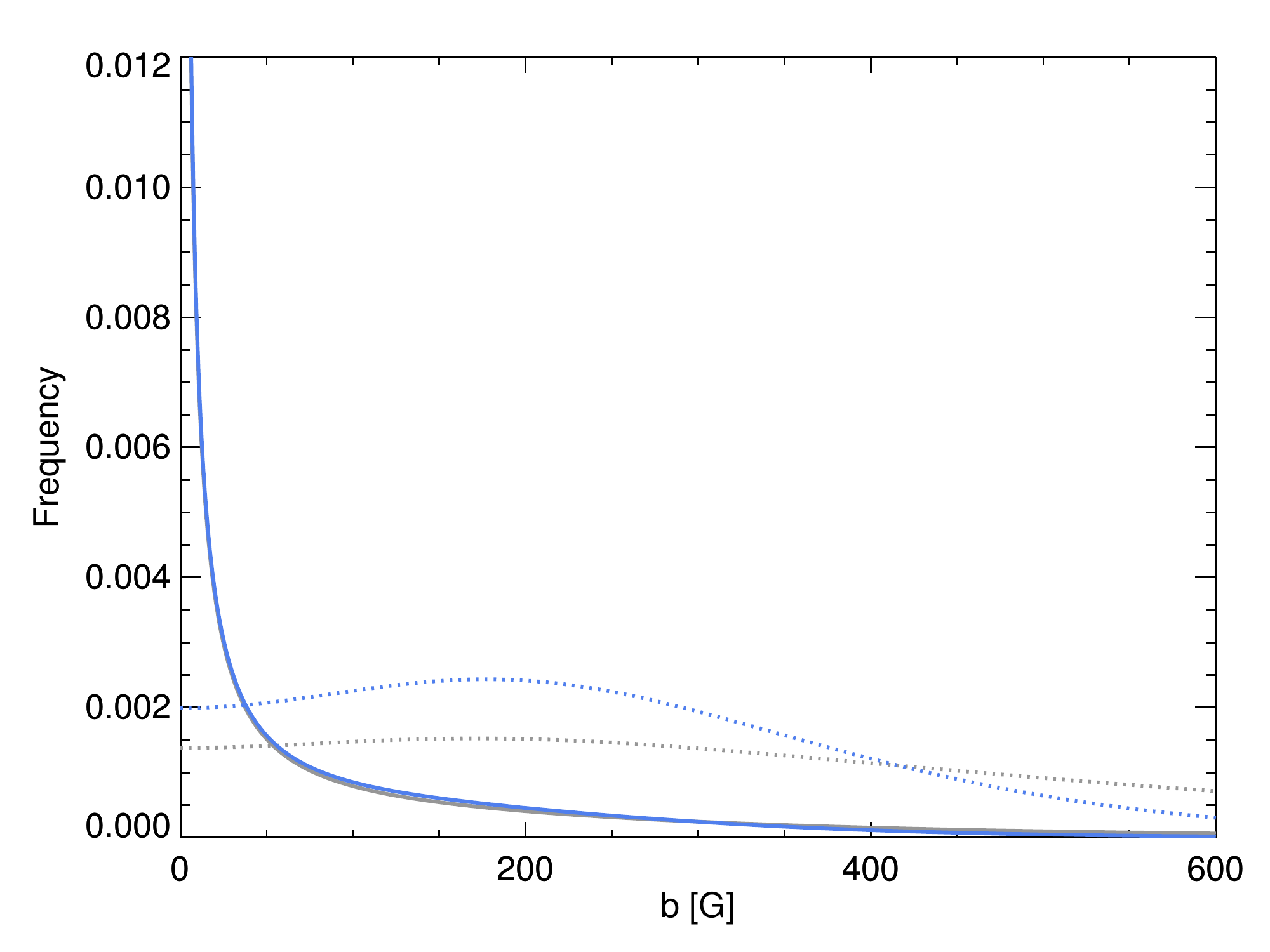}
\caption{Likelihood (dotted line) and posterior (solid line) for the parameter $b$. The grey lines show the
results when only the data of \protect\cite{Leone+14} is taken into account. When the data of \protect\cite{Jordan+12} is added, the
curves are updated to the blue lines. 
The posterior is equal to the likelihood multiplied by the Jeffrey's 
prior (see equation~[\ref{eq09}]).}
\label{fig:posterior_b}
\end{figure}

We now make the simplification of assuming that all observations are statistically independent. In other words, 
observing an object tells you nothing about the magnetic field of another one. Repeated observations of the 
same star are modeled by using the same $B_{\parallel i}$ unless
the observations are obtained in too different epochs. In this case, they are considered to be different
objects because of the potential variability of the source.
Strictly speaking, the assumption of independence is surely not correct for different observations of the
same star at different epochs and one should include the covariance between them. 
However, we make this simplification for the sake of obtaining a simpler analytical final result
and because this covariance is difficult to model and estimate. If the previous conditions hold,
we can factorize the likelihood and the priors so that the posterior reads
\begin{equation}\label{eq05}
p(B_{\parallel 1}, \dots, B_{\parallel N}|\{D_i\})\propto \prod_{i=1}^{N_{\rm obs}} p(D_i|B_{\parallel i})p(B_{\parallel i}),
\end{equation}
where $\{D_i\}$ is the set of all the observed Stokes profiles averaged over the stellar surface,
$p(D_i|B_{\parallel i})\equiv {\cal L}_{B_{\parallel i}}$, 
and $p(B_\parallel)$ is a Gaussian with mean 0 and variance $b^2$.
Instead of using a fixed $b$, we follow a hierarchical Bayesian approach and we introduce $b$ 
into the Bayesian inference scheme \citep{Gregory05}, so that
\begin{equation}\label{eq05b}
p(B_{\parallel 1}, \dots, B_{\parallel N},b|\{D_i\})\propto \prod_{i=1}^{N_{\rm obs}} p(D_i|B_{\parallel i})p(B_{\parallel i},b).
\end{equation}
The $b$ parameters
is then termed \emph{hyperparameter}, since it is a parameter of the prior. Note that is does not
affect the likelihood. The joint prior for $B_\parallel$ and $b$ can be factorized as
\begin{equation}\label{eq06}
p(B_\parallel, b)=p(B_\parallel|b)p(b),
\end{equation}
so that the marginal prior for $B_\parallel$ could be obtained by marginalizing $b$
(i.e., integrating out)
\begin{equation}\label{eq07}
p(B_\parallel)=\int p(B_\parallel|b)p(b) \mathrm{d}b.
\end{equation}
As discussed above, we consider $p(B_\parallel|b) = {\cal N}(0, b^2)$.
Integrating equation (\ref{eq05}) over all the $B_{\parallel i}$ variables (using equation
[\ref{eq03}]), the marginalized posterior for $b$ is obtained, resulting in
\begin{equation}
p(b|\{D_i\}) \propto p(\{D_i\}|b)\;p(b),
\label{eq09}
\end{equation}
where $p(\{D_i\}|b)\equiv {\cal L}_b$ is given by 
\begin{equation}
\begin{split}\label{eq08}
{\cal L}_b=&\prod_{i=1}^{N_\mathrm{obs}} \int \mathrm{d}B_{\parallel i}\; p(D_i|B_{\parallel i}) p(B_{\parallel i}|b) 
\\
=&
\left[\prod_{i=1}^{N_{\rm obs}} \left[\prod_{s=1}^S 
\frac{1}{(2\pi\sigma_{V}^2)^{L_{s_i}/2}} \right]  \frac{1}{(1+2C_{1i}b^2)^{1/2}}
 \right]
\\
&\times \exp\left\{ -\sum_{i=1}^{N_{\rm obs}}\frac{\hat{B_\parallel}_i^2 C_{1i}}{1+2C_{1i}b^2} 
-\sum_{i=1}^{N_{\rm obs}} C_{0i}   \right\},
\end{split}
\end{equation}
while $p(b)$ is the prior to be chosen over the hyperparameter $b$.
The previous expression can be considered as a multiobject and multiline approach
to magnetic field detection.

\begin{figure}
\includegraphics[width=\columnwidth]{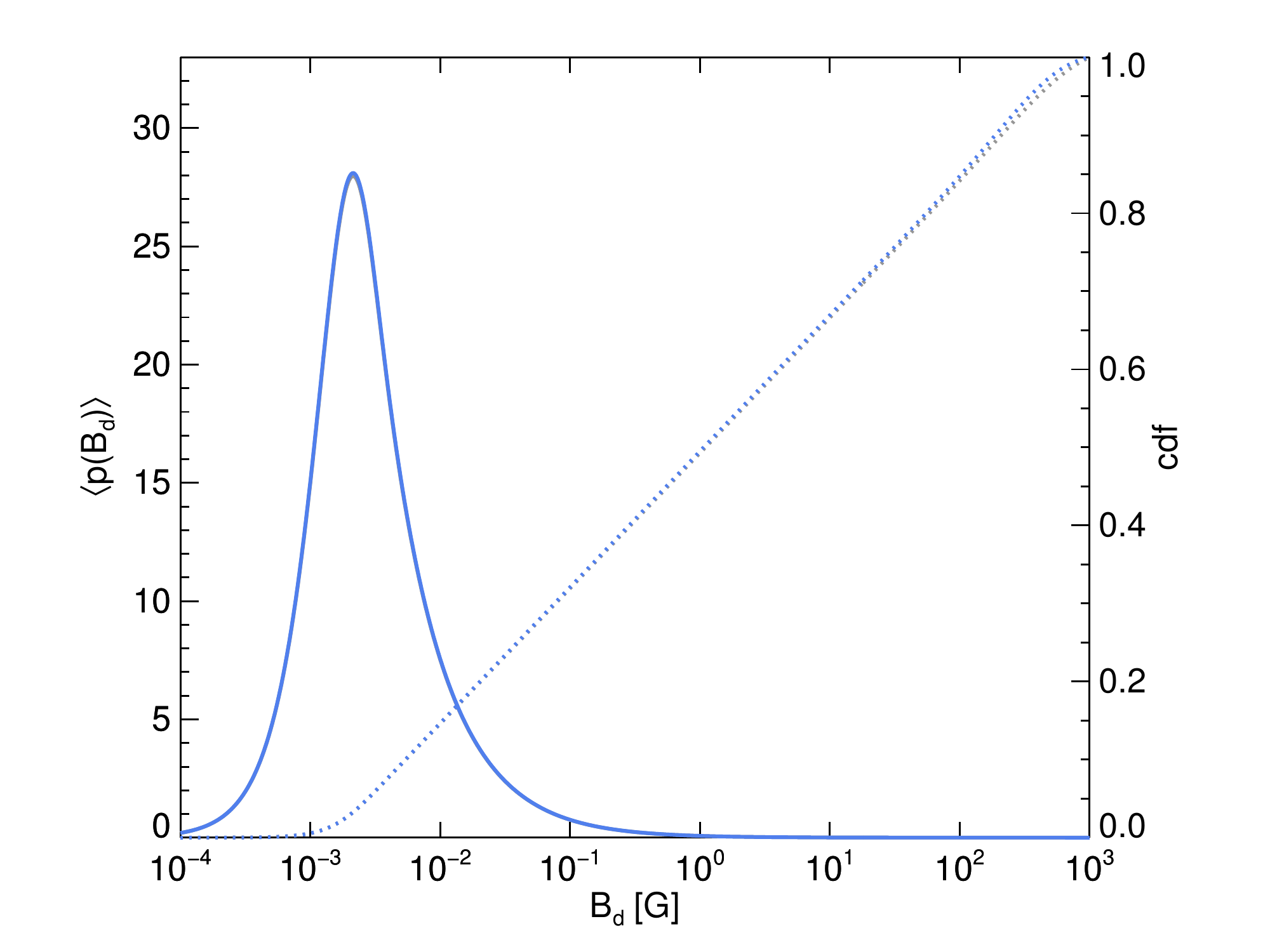}
\caption{Distribution for the magnetic field strength that emerge from the observations, under
the assumption that all CSPN are samples of a unique distribution. The probability
distribution function are shown in solid lines, while the cumulative distribution is
shown in dotted lines. Grey curves correspond to data of \protect\cite{Leone+14} alone, while
blue curves are computed adding the data of \protect\cite{Jordan+12}. Only the blue curve
is conspicuous because the grey and blue curves overlap.}
\label{fig:posterior_bmod}
\end{figure}

We do not have any a priori information on $b$, which is a scale variable.
A convenient uninformative prior in these cases is the Jeffreys prior $p(b)\propto 1/b$
because it distributes probability uniformly in the logarithm: it is as likely to 
find $b$ between 0.1 and 1~G, as between 10 and 100~G
\citep{Jeffreys68, Jaynes03}.
This prior is improper (its integral is not finite) and here, it yields an improper posterior too.
For that reason, a cutoff $b_\mathrm{min}$ has to be chosen.
For any informative data set (peaked ${\cal L}_b$), the value of the cutoff becomes irrelevant
but for, perhaps, ridiculously small values; for less informative (i.e., flatter) ${\cal L}_b$, 
we must check the sensitivity of the results to the value of $b_\mathrm{min}$.

Figure \ref{fig:posterior_b} shows, in grey, ${\cal L}_b$ (dotted line) and the posterior $p(b|\{D_i\})$ using a Jeffreys' prior (solid
line) for the CSPN data compiled in \cite{Leone+14} and summarized in table~\ref{tab:table}.
A characteristic of the likelihood displayed in equation (\ref{eq08}) is that it only
depends on the maximum likelihood estimation of the longitudinal field, $\hat{B}_\parallel$ and
of the coefficients $C_1$ (the coefficients $C_0$ represent just a scaling of the
likelihood and are of no importance for our purpose). According to equation (\ref{eqa9}), $C_1$ can be obtained from the
uncertainty in the estimation of $\hat{B}_\parallel$, as computed by \cite{Martinez+12}. Therefore,
we can make use of the recent results of \cite{Jordan+12} and use their estimations to upgrade our
observations. Note that, since their results represent a longitudinal magnetic field averaged on the stellar surface, 
they have to be multiplied by 4 to account for the dipolar dilution \citep{Martinez+12}.
According to Fig. \ref{fig:posterior_b}, the addition of the observations of \cite{Jordan+12}
barely modifies the posterior distribution $p(b|\{D_i\})$. The results point to a very small value
of $b$.

From the posterior of the hyperparameter, we derive the posterior probability 
of the longitudinal field for all the sources as
\begin{equation}\label{eq10}
\langle p(B_\parallel) \rangle=\int \mathrm{d}b \; p(B_\parallel|b) \; p(b|\{D_i \}).
\end{equation}
This distribution depends on the cut-off $b_\mathrm{min}$, but 
only weakly for $b_\mathrm{min}$ ranging between $10^{-6}$-$10^{-1}$~G. Although we do not
display the figure, choosing $b_\mathrm{min}=1$~mG, $B_\parallel$ is below
$\sim$250~G in absolute value with 95\% probability.

More interesting is to consider the distribution of $B_d$, the magnetic field
strength of the dipole,  which can be obtained from the distribution of $B_\parallel$
assuming that the dipole inclination angle of the different CSPN is randomly (isotropically)
oriented in space, which seems reasonable since there is no 
clear evidence for a preferential orientation of PNe themselves \citep{Corradi+98}.
This computation is not possible in the general case in which $B_\parallel$ is
interpreted as an average longitudinal field over the stellar surface.
Then, a-priori, $B_d$ follows a Maxwellian distribution:
\begin{equation}
p(B_d|b) =\frac{4\pi}{(2\pi b^2)^{3/2}}B_d^2\exp\left( -\frac{B_d^2}{2 b^2} \right).
\end{equation}
Proceeding as above we derive the distribution of dipolar magnetic field
strength from all the observed sources as
\begin{equation}\label{eq11}
\langle p(B_d) \rangle=\int \mathrm{d}b \; p(B_d|b) \; p(b|\{D_i \}).
\end{equation}
It is interesting to compare this equation with Eq. (7) of \cite{kolenberg_bagnulo09} and 
point out the differences. Although the two formalism assume an isotropic distribution
of fields, our approach is more general for the following reasons:
i) we use a fully Bayesian approach in which we model the observed Stokes profiles to 
correctly propagate all uncertainties in the observations to the magnetic field
distribution, while \cite{kolenberg_bagnulo09} use the maximum likelihood estimation of the $B_\parallel$ and ii)
we take into account that each source can have a different magnetic field
strength and that it is known with imprecision.

Figure \ref{fig:posterior_bmod} shows the corresponding
probability density function (solid) and the cumulative distribution (dotted) for the
data of \cite{Leone+14} alone and what happens when we add the
data of \cite{Jordan+12}.
From this, the dipolar magnetic field strength is larger than 3 mG and smaller than
400 G with 95\% probability. 
Note that this result is independent of $b_\mathrm{min}$,
since equation~(\ref{eq11}) is not divergent for $B_d=0$, unlike 
equation~(\ref{eq10}). 


\section{Discussion and conclusions}
\begin{figure}
\includegraphics[width=\columnwidth]{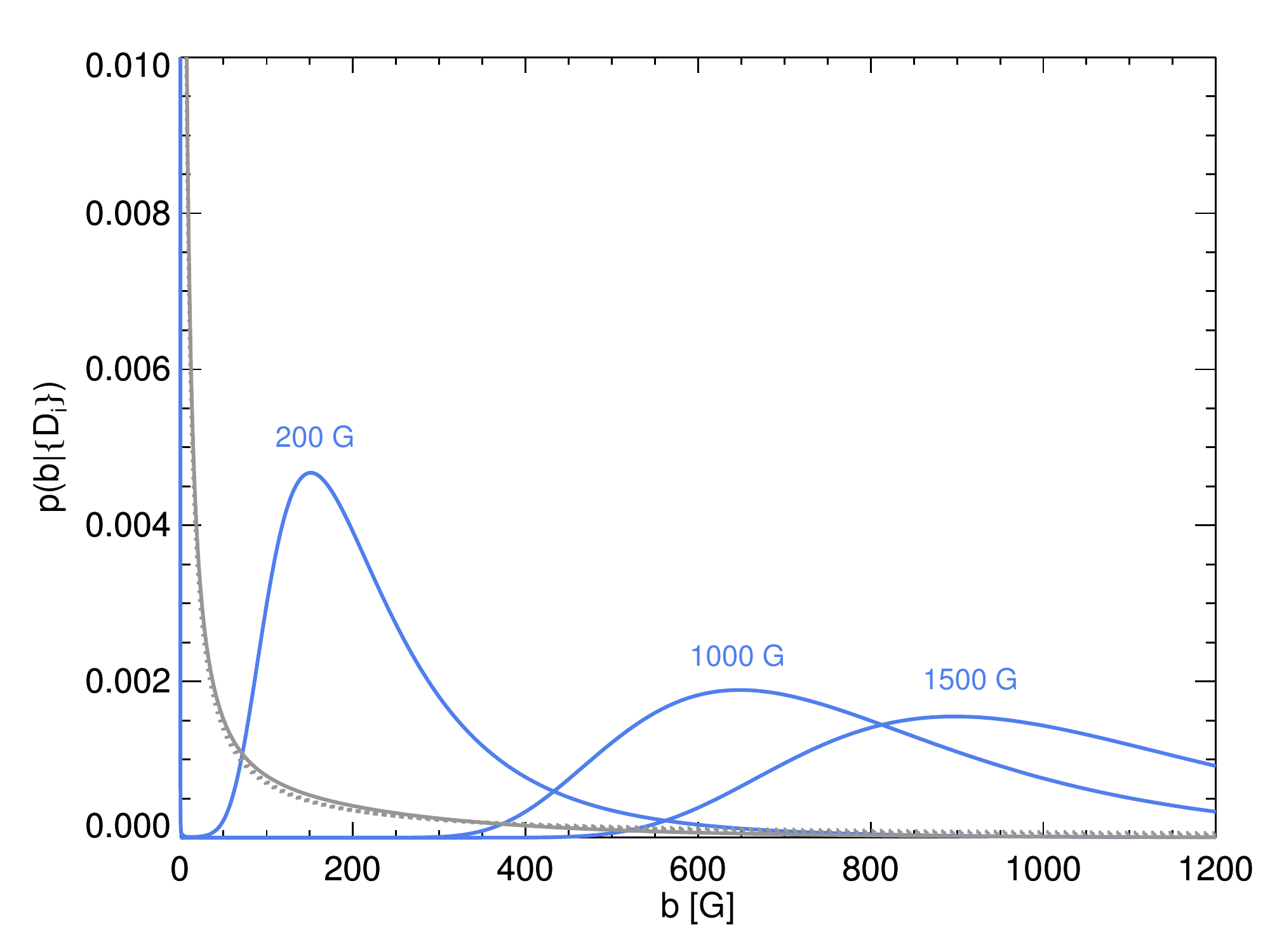}
\caption{Illustration of how adding two new objects with clear 
magnetic field detections of 200~G, 1000~G, or 1500~G, would lead
to a rapid change of the posterior (blue lines). 
By contrast, if had observed just four objects out of the total sample,
the posterior would be far less informative (dotted lines for three
random realizations of four elements, superposed), than the complete sample to date
(solid grey line).} 
\label{fig:posterior_b_newmeasurements}
\end{figure}

An important property of ${\cal L}_b$ is its ability to concentrate for large $N_\mathrm{obs}$.
Consider, for example, an ensemble of objects with similar spectra 
that are observed with identical uncertainties $\sigma_V$
(hence, they have similar $C_1$);
they only differ in their magnetic field strength (i.e., $\hat{B}_{\parallel i}$).
Then, the maximum of ${\cal L}_b$ (or $\log{\cal L}_b$) in equation~(\ref{eq08}) 
will lie at 
\begin{equation}
\begin{array}{ll}
b_{\rm ML}=0  & \mbox{if } \langle \hat{B}^2\rangle < \frac{1}{2C_1} \\
b_{\rm ML}=\frac{\sqrt{2C_1 \langle \hat{B}^2\rangle}-1}{2C_1} & \mbox{if } \langle \hat{B}^2\rangle \geq \frac{1}{2C_1},
\end{array}
\end{equation}
where $\langle \hat{B}^2\rangle = \sum_i B_{\parallel_i}^2/N_\mathrm{obs}$.
In the first case, the observations are inadequate (too noisy, and therefore, $C_1$ too large),
for the estimated $\hat{B}_i$ to be informative.
${\cal L}_b$ is relatively flat near $b=0$ and 
when multiplied by the Jeffreys prior, the posterior clearly diverges towards smaller values,
always reaching its maximum a-posteriori (MAP) value at the cut-off, $b_\mathrm{MAP}=b_{\rm min}$, which is not very helpful.
An illustration of this situation is given by one of the dotted lines 
in figure~\ref{fig:posterior_b_newmeasurements}, which corresponds to an incomplete subset of 
just four observed objects from table~\ref{tab:table}.
The two other dotted lines correspond to different subsets with the same number of objects.
In those other cases, ${\cal L}_b$ has a maximum at some non-zero value $b_{\rm ML}$,
which is noticeable also in the posterior $p(b|{D_i})$.
But they are notably wider (spanning several hundreds of gauss) than the 
likelihood obtained from the full set of observed objects (gray solid line).
In fact, it can be shown that for large values of $N_{\rm obj}$, 
the width of ${\cal L}_b$ converges to 0 as $N_{\rm obj}^{-1/2}$. 
Therefore, ${\cal L}_b$ becomes more and more informative on the value of $b$ 
as we accumulate observations ($N_\mathrm{obs}\rightarrow\infty$),
even if the individual observations are too noisy to get a clear detection
(cf. solid gray line in Figure~\ref{fig:posterior_b_newmeasurements}).
In a sense, this is a generalization of {\em multiline addition techniques} 
in which the collective contribution of many individual spectral lines from an object is
used to increase the signal-to-noise ratio of the polarization pattern
\citep{Semel96, Donati+97, Martinez+08}.
Here, the collective contribution of many different objects contribute to constrain  
the statistical distribution of fields. 

It is also interesting to see how a few {\em better} observations may affect  
the estimates. For example, consider that we observed two objects with 
well-defined $\hat{B}_\parallel$, but with a much better precision than up to now
(say, $\sigma_V$ is a factor 100 better), i.e., in case we had a clear detection for some 
(perhaps, different) object. Figure~\ref{fig:posterior_b_newmeasurements} shows how $p(b|\{D_i\})$
change from the original (solid grey line) for three different values of $\hat{B}_\parallel$.
Likewise, we also display in dotted grey line what happens if we only take into account 4 of the available
observations. This demonstrates that, although there is not a single detection of magnetic
fields, adding more objects produces a collapse of the posterior.

We have introduced two major simplifications in the method which are not essential.
The first one regards noise estimation. We have assumed that the variance 
of the observed quantities ($\sigma_V$ and $\sigma_{I'}$) was known for the derivation
the likelihood functions (see Appendix), and that they are independent of wavelength.
Their actual values have been independently estimated 
from the intensity fluctuations in the continuum windows between spectral lines.
Estimating the unknown variance of a set of measurements is a fundamental problem
in Bayesian theory which can be done independently (like here) or 
consistently within the Bayesian analysis itself by assigning (e.g., non-informative, 
Jeffreys) priors to $\sigma_V$ and $\sigma_{I'}$, and marginalizing these parameters.
We have not pursued this more general approach here to keep our main argument simple.  

Secondly, the approximation in equation~(\ref{eq03}) has allowed the analytical derivation
of equation~(\ref{eq08}). This approximation is accurate beyond the strict 
limits stated above. When the more general equation~(\ref{eq02}) is required, 
the integrals in equation~(\ref{eq08}) have to be performed numerically using
Markov Chain Monte Carlo methods or alike, depending on the dimensionality of the problem.

An important characteristic of the analysis presented is that it naturally
allows integration of new data to improve the magnetic field estimates. We have
already shown how things change when the data of \cite{Jordan+12} is added to
the observations of \cite{Leone+14}. From the analysis of all the available observations so far we have 
obtained the upper-limit of 400 G.

Our analysis here suggests several ways in which such estimates can be improved.
As shown above, the mere addition of new observations helps constraining 
the magnetic field distribution, even when no clear detection in the individual 
objects is possible.
The constraints on the global distribution can be even stronger when either
the new observations have a better (lower) noise level, or if clear
detection on individual objects is achieved. 

The analysis presented in this paper is not limited to any particular spectral range, provided
that the weak-field approximation holds. Observations at different spectral windows can be straightforwardly included in the analysis after
computing their corresponding $C_{1}$, $\hat{B}_\parallel$, and $\sigma_V$ values
(although in some cases it might be advisable to consider the general expression
for the likelihood (Eq.~(\ref{eq02})).
Given that the amplitude of the Zeeman Stokes $V$ scales with $\lambda_0$ \citep[see, e.g.,][]{landi_landolfi04}, spectropolarimetry of the Paschen and Brackett series 
should be favoured.

The joint analysis of linear and circular polarization would impose
stronger constrains on the magnetic field distribution under the, very likely,
assumption of isotropic distribution of fields for the observed objects. 
The Zeeman effect generates linear polarization patterns
which are usually smaller than those of circular polarization and that are, in the weak-field
approximation, proportional to the square of the transversal component
$B_\perp^2$ of the magnetic field to the LOS \citep[e.g.,][]{Landi92, L4, Martinez+12}.
However, we understand that the detection of linear polarization is extremely improbable, 
given that no reliable detection of circular polarization has been achieved so far.

Additionally, magnetic alignment of dust grains creates linear polarization in the continuum 
from which we may infer the presence of a magnetic field \citep[e.g.,][]{DavisGreenstein51, Lazarian07}. 
The information thus obtained is not quantitative and cannot be directly included
in our formalism. Yet, it may provide important general and symmetry constraints 
and bounds that could be implemented within the Bayesian methodology.

Finally, it is clear that the approach presented here can be applied to other sets of objects 
(e.g., white dwarfs, \ldots) once they can be assumed to belong to the same magnetic class, i.e., they are
all characterized by the same statistical distribution of fields.

\acknowledgements
The authors are grateful to Stefano Bagnulo for some interesting suggestions and for a careful review of the paper.
We also thank C. Gonz\'alez for useful discussions.
Financial support by the Spanish Ministry of Economy and Competitiveness and the
European FEDER Fund through projects AYA2010-18029 (AAR, MJMG, and RMS), AYA2012-35330
(RLMC), Consolider-Ingenio 2010 CSD2009-00038 (AAR and RMS), and Ram\'on y Cajal
fellowship program (AAR), is greatly acknowledged.

\appendix
\section{Computation of the likelihood}
We assume that the observed circular polarization flux is given by equation~(\ref{eq01}) 
but that it is corrupted by a normaly distributed noise $e_V\sim {\cal N}(0, \sigma_V^2)$ 
with zero mean and variance $\sigma_V^2$:
\begin{equation}\label{eqa1}
F_V=-\alpha B_\parallel {\cal F}_I' + e_V;
\end{equation}
the flux derivative, in turn, is corrupted by a noise $e_{I'} \sim {\cal N}(0, \sigma_{I'}^2)$:
 \begin{equation}\label{eqa2}
F_I'= {\cal F}_I' + e_{I'}.
\end{equation}
Assuming that both sources of error are independent, then the joint likelihood of the data is
\begin{equation}\label{eqa3}
\begin{split}
p(F_V, F_{I'}|B_\parallel, {\cal F}_I', \sigma_V, \sigma_{I'})
&=p(F_V|B_\parallel, {\cal F}_I', \sigma_V)p(F_I|{\cal F}_I', \sigma_{I'})
\\
&=\frac{1}{2\pi\sigma_V\sigma_{I'}}\exp\left\{ -\frac{(F_V+\alpha B_\parallel {\cal F}_I')^2}{2\sigma_V^2}\right\}
\exp\left\{-\frac{(F'_I-{\cal F}_I')^2}{2\sigma_{I'}^2}  \right\}.
\end{split}
\end{equation}
The {\em true} value ${\cal F}_I'$ is treated as a nuisance parameter that we may integrate out for some uniform vague prior 
\begin{equation}\label{eqa4}
p(F_V, F_{I'}|B_\parallel, \sigma_V, \sigma_{I'})=\frac{1}{\sqrt{2\pi}}
\frac{1}{\sqrt{\sigma_V^2+\alpha^2 B_\parallel^2 \sigma_{I'}^2}}
\exp\left\{ -\frac{(F_V+\alpha B_\parallel F'_I)^2}{2(\sigma_V^2+\alpha^2 B_\parallel^2 \sigma_{I'}^2)} \right\}.
\end{equation}
Assuming statistical independence for the $L$ wavelengths, and constant $\sigma_V$ and $\sigma_{I'}$ accross the spectral line,
\begin{equation}\label{eqa5}
p(\{F_V(\lambda_\ell), F'_{I}(\lambda_\ell)\}|B_\parallel, \sigma_V, \sigma_{I'})=
\frac{1}{[2\pi(\sigma^2_V+\alpha^2 B_\parallel^2 \sigma_{I'}^2)]^{L/2}}
\exp\left\{ -\frac{c_1(B_\parallel-\hat{b}_\parallel)^2+c_0}{2(\sigma^2_V+\alpha^2 B_\parallel^2 \sigma_{I'}^2)} \right\},
\end{equation}
where, introducing the notation
$\langle F_V^2\rangle=\sum_\ell F_V(\lambda_\ell)^2$, 
$\langle F_V F'_{I}\rangle=-\alpha\sum_\ell F_V(\lambda_\ell)F'_I(\lambda_\ell)$, 
$\langle {F'_{I}}^2\rangle=\alpha^2\sum_\ell F'_I(\lambda_\ell)^2$,
then, 
\begin{equation*}
c_1=\langle {F'_I}^2\rangle, \quad c_0=\langle F_V^2\rangle-c_1\hat{b}_\parallel^2, \quad \mathrm{and}\quad \hat{b}_\parallel=\frac{\langle F'_I F_V \rangle}{\langle {F'_I}^2 \rangle}.
\end{equation*} 
Note that $c_{0, 1}$, and $\hat{b}_\parallel$ implicitly depend on the spectral line considered 
and also explicitly through the effective Land\'e factor within $\alpha$.
Extending the argument to all the $S$ spectral lines 
\begin{equation}\label{eqa6}
{\cal L}_{B_\parallel}\equiv
p(\{   F_V(\lambda_{\ell,s}), F'_{I}(\lambda_{\ell,s})    \}|B_\parallel, \sigma_V, \sigma_{I'})=
\left[\prod_{s=1}^{S}\frac{1}{[2\pi( \sigma^2_V + \alpha_s^2 B_\parallel^2 \sigma_{I'}^2)]^{(L_s/2)}}\right]
\exp\left\{ -\sum_s\frac{c_1(B_\parallel-\hat{B}_\parallel)^2+c_0}{2(\sigma^2_V+\alpha_s^2 B_\parallel^2 \sigma_{I'}^2)} \right\},
\end{equation}
where $\lambda_{\ell,s}$ represents the wavelength point $\ell$ of line $s$.
In equation~(\ref{eqa6}), $\sigma_V$ and $\sigma_{I'}$ are, in principle, 
different for the each spectral line; they are estimated from their adjacent continuum 
(for clarity, we do not write the subscript $s$).
To second order on $\zeta=B_\parallel \sigma_{I'}/\sigma_V$,
\begin{equation}\label{eqa7}
{\cal L}_{B_\parallel}=
\left[\prod_{s=1}^{S}\frac{1}{[2\pi \sigma^2_V]^{(L_s/2)}}\right]
\exp\left\{ -C_1(B_\parallel-\hat{B}_\parallel)^2-C_0 \right\} \;+\mathrm{O}(\zeta^2)
\end{equation}
where 
\begin{equation}\label{eqa8}
C_1=\langle\!\langle {F'_I}^2\rangle\!\rangle, 
\quad 
C_0=\langle\!\langle F_V^2\rangle\!\rangle - C_1 \hat{B}_\parallel^2,
\quad \mathrm{and}\quad 
\hat{B}_\parallel=\frac{\langle\!\langle F'_I F_V \rangle\!\rangle}{\langle\!\langle {F'_I}^2 \rangle\!\rangle},
\end{equation}
with 
$\langle\!\langle F_V^2\rangle\!\rangle=\sum_{\ell s} F_V(\lambda_{\ell,s})^2 /(2\sigma_V^2)$; 
$\langle\!\langle F_V F'_{I}\rangle\!\rangle=-\sum_s[\alpha_s/(2\sigma_V^2)]\sum_\ell F_V(\lambda_{\ell,s})F'_I(\lambda_{\ell,s}) $; and 
$\langle\!\langle {F'_{I}}^2\rangle\!\rangle=\sum_s[\alpha_s^2/(2\sigma_V^2)]\sum_\ell F'_I(\lambda_{\ell,s})^2$. Interestingly,
the $C_1$ coefficient can be related to the error bar of $\hat{B}_\parallel$ as shown by \cite{Martinez+12}
\begin{equation}\label{eqa9}
C_1 = \frac{1}{2 \sigma_{B_\parallel}^2}.
\end{equation}
Therefore,
all the ingredients to carry out our calculations are readly available from any work that tabulates the maximum-likelihood
estimation of the longitudinal field and its associated error bar.


\end{document}